\documentclass[a4paper,11pt]{article}
\begin{document}
\title{Mechanism of irreversibility in a many-body systems}
\author{V.M. Somsikov}
\date{\it{Laboratory of Physics of the geoheliocosmic relation,\\
Institute of Ionosphere, Almaty, 480020, Kazakhstan\\E-mail:
nes@kaznet.kz}} \maketitle
\begin{abstract}
The mechanism of irreversible dynamics in the mixing systems is
constructed in the frames of the classical mechanics laws. The
offered mechanism can be found only within the framework of the
generalized Hamilton's formalism. The generalized formalism is
created by expansion of the canonical Hamilton's formalism to the
open systems. A formula, which expresses the entropy through the
work of subsystems interaction forces was obtained. The essential
link between thermodynamics and classical mechanics was
established.
\end{abstract}.

\section{Introduction}

Irreversibility is a main difficulty in aspiration of linking
classical mechanics with thermodynamics [1-4]. A first attempt to
overcome this difficulty has been done by Boltzmann. He found that
many-body systems should be equilibrated. But for obtaining this
result, Boltzmann had used probabilistic principles. Therefore the
difficulty was not overcome. Since Boltzmann and till now the
attempts to overcome irreversibility problem do not stop. For this
purpose practically all areas of physics are used: statistical
physics, kinetics, classical and quantum mechanics, nonequilibrium
dynamics and so on.

In a basis of statistical physics and kinetics probabilistic laws
are used. They allow to describe effectively collective properties
of systems, having refused from an inconceivable task of
calculation of dynamics of each element. So, for example, the
method of Gibbs' microcanonical ensembles consists in splitting
equilibrium system to subsystems and studying them, basing on
probabilistic laws in the assumption of ergodicity hypotheses
performance [2, 4-6]. The statistical theory of equilibrium
systems was created having excluded with the help of ergodicity
hypotheses the interaction of subsystems and dependences of their
statistical distribution from the entry conditions [2]. But a
similar sort of a hypotheses and probabilistic principles for a
substantiation of irreversibility is unacceptable.

Near ten years ago the method of non-extensive thermodynamics,
applicable for the analysis of stationary nonequilibrium systems
has been arisen [7]. It allows to determine function of
distribution of nonequilibrium systems and to study connection
between parameters of thermodynamic and mechanic [8]. The certain
successes in studying nonequilibrium systems within the framework
of the statistical theory of open systems have been achieved [9].
The feature of the offered approach is taking into consideration
the structure of the continuous environment at all levels of the
description. But this method is inapplicable for the solving of
irreversibility problem because it is also based on probabilistic
laws.

Attempts to prove thermodynamics on the basis of strict methods of
classical mechanics were undertaken also. It has been proved by
Liouville, that only those systems of many bodies, which can be
splittied on systems with one degree of freedom by transformations
of independent variables, are integrable [10, 11]. I.e., the
system is integrable, when it is possible to exclude interactions
between its elements. Poincare also has proved the theorem,
according to which the dynamic systems in most cases are not
integrable because of impossibility of exception of forces acting
between subsystems [10]. But if potentiality of the forces between
elements means potentiality of forces between systems of these
elements, as it follows from a formalism of Hamilton [12], the
problem will be reduced to independent integrable systems with one
degree of freedom. The contradiction is obviously. On the one
hand, it is proved, that the class of integrable systems is very
narrow. On the other hand, potentiality of interacting forces
should provide an opportunity to transform the natural systems to
the integrable systems [12].

The discovering of the deterministic chaos was the reason of
creating of the stochastic dynamics. Stochastic dynamics is
constructed on the basis of laws and principles of classical
mechanics [13-15]. Basing on the methods of stochastic dynamics,
the connection between entropy and Lyapunov's exponents was
established. Performance of the mixing property for dynamic
systems was proved. Mixing provides a randomness of dynamics and
decay correlations. Basing on the mixing properties the modern
mechanism of irreversibility has been offered [1, 10, 13-15]. But
this mechanism has an insuperable barrier. It is explanation of
the nature of "coarse-grain" of the phase-space. But Poincare has
proved impossibility of its explanation within the framework of
canonical formalism of classical mechanics. Thus the
irreversibility problem was reduced to the problem of
"coarse-grain" of the phase-space [13-15].

I. Prigogine has guessed that difficulties of the solution of a
problem of irreversibility may be connected with limitation of a
canonical formalism of classical mechanics for the description of
real systems. Indeed, this formalism is applicable to conservative
systems while irreversibility is caused by their openness [10].
This assumption has defined our approach to researches of a
problem of irreversibility.

The goal of this work is investigation about how the mixing
creates irreversibility and what is the nature of the link between
classical mechanics and thermodynamics. We will do it by extending
Hamilton's formalism so that it will be applicable to analyzing of
the open systems.

Our investigation is based on the following method. A conservative
system of hard disks and potentially interacting elements, which
is not in equilibrium, is prepared. This system is then split into
small subsystems that are accepted as being in equilibrium. The
subsystem dynamics under condition of their interactions is
analyzed on the basis of classical mechanical laws. It allows us
to reduce a problem of the description of dynamics of systems
elements to a problem of the dynamics of interacting equilibrium
subsystems.

The researches were constructed in the following way[16-18, 20,
21]. The generalized Liouville equations was obtained. These
equations are applicable either for the description of systems of
hard disks, and of open systems. Based on the generalized
Liouville equation the necessary condition for occurrence of
irreversible dynamics has been obtained, and an opportunity of
existence of irreversibly dynamics in a disk system has been
proved.

The dynamics of a hard-disks system was studied. The equation of
motion for them has been obtained. Based on this equation the
non-potentiality of the interaction forces of disks and their
subsystems was established.

The mechanism of irreversibility of hard disks has been
generalized to the systems consisting of potentially interacting
elements. It has been made basing on the equation for energy
exchange between interacting subsystems. This equation has been
obtained from the law of conservation of energy. The
non-potentiality of forces acting between the subsystems which
consists of the potentially interacting elements has been
established.

The connection between classical mechanics and the first
thermodynamics law was found. The formula for entropy production
has been obtained.

\section{The general Liouville equation}

The approach to the solution of a problem of irreversibility
offered here is based on the generalized Liouville equation. Let
us to find out than this equation is differs from the canonical
Liouville equation, which lay in the bases of canonical Hamilton's
formalism?

The canonical Liouville equation is convenient for systems under
conditions of short enough times when it is possible to neglect an
exchange of energy between subsystems [2, 3]. Moreover, it is
applicable for the description of only potentially interacting
systems. But in the nonequilibrium hard disks system, the
interaction forces of subsystems are non-potentially [17]. Thus,
the description of dynamics of nonequilibrium hard disks systems
within the framework of the canonical equation with these
restrictions is impossible. These restrictions for the generalized
Liouville equation are absent.

The generalized Liouville equation for subsystems was obtained by
the next way [17, 18]. We took a closed nonequilibrium system,
which consist of $N$ elements. Divide this system into $R$
equilibrium subsystems. Then we select one of subsystem, which we
call $m$-subsystems. With the help of D'Alambert equation, basing
on variational method, the generalizing Lagrange, Hamilton
equations were obtained for $m$-subsystem. Basing on these
equations, the generalized Liouville equation was obtained within
the framework of laws of classical mechanics. This equation can be
written so [16]:
\begin{equation}
{\frac{df_m}{dt}=-f_m\sum\limits_{k=1}^L
\frac{\partial}{\partial{\vec{p}_k}}F_{k}^m} \label{eqn1}
\end{equation}

Here $f_m=f_m({\vec{r}}_k,{\vec{p}}_k,t)$ is a normalized
distribution function of subsystem disks; ${F^{m}_{k}}$ is
external force acting on $k$- disks of $m$-subsystem from outside,
$F_{k}^m=\sum\limits_{s=1}^{N-L}F_{ks}^m$; ${k=1,2...L}$ are disks
of $m$-subsystem; the $s= 1, 2, 3,..., N-L$ are external disk
acting on $k$-disk of $m$-subsystem; $m=1, 2, 3,...R$;
${\vec{p}}_k$ and ${\vec{r}}_k$ are momentous and coordinates for
$m$-subsystem disks consequently; $t$ is a time.

The right hand side of the eq. (1) plays the role of the
collisions integral. It is equal to zero, when the forces are
potential, and when the system is in equilibrium state. The right
hand side can be found from the equations of motion for elements
of system.

The important step on the way to the generalize Liouville equation
were failure from a requirement of potentiality of forces between
subsystems. Therefore this equation is fair for anyone opens and
nonholonomic systems. It does applicable this equation for the
study of irreversibility [12, 18].

Let us remark, that the similar form of the generalized Liouville
equation can be obtained if accepted forces is dissipative [19].
But in this case the generality of the obtained equation will be
lost. It will be so because the presence of the dissipative forces
is equivalent to irreversibility. Moreover, acceptance of
condition dissipative forces essentially narrows area of
applicability of this equation. Indeed, forces of interaction of
disks though are non-potential, but they and nondissipative [12].
Let us call further the forces between subsystems as "generalized
forces".

 Let us consider the important properties of dynamics that
directly follow from generalized Liouville equation. These
properties can be obtained by the analysis of character of
interrelation of dynamics of selected subsystems with dynamics of
system as a whole. They are caused by preservation of phase-space
for the full system [18].

As the equality, ${\sum\limits_{m=1}^R{\sum\limits_{k=1}^L
F_{k}^m=\sum\limits_{m=1}^{R}F_m =0}}$, is carried out, the next
equation for the full system Lagrangian, $L_R$, will have a place:
${\frac{d}{dt}\frac{\partial{L_R}}{\partial{v_k}}-
\frac{\partial{L_R}}{\partial{r_k}}=0}$ and the appropriate
Liouville equation:
${\frac{\partial{f_R}}{\partial{t}}+{v_k}\frac{\partial{f_R}}
{\partial{r_k}}+\dot{p_k}\frac{\partial{f_R}}{\partial{p_k}}=0}$.
Here $f_R$ is a distribution function corresponds to the full
system; $v_k$ is a velocity of $k$-disk. The full system is
conservative. Therefore, we have: ${\sum\limits_{m=1}^R
divJ_m=0}$. Here, $J_m=(\dot{\vec{r}_k},{\dot{\vec{p}_k}})$ is a
generalized current vector of the $m$-subsystem in a phase space.
This expression is equivalent to the next equality:
${\frac{d}{dt}(\sum\limits_{m=1}^{R}\ln{f_m})}=
\frac{d}{dt}(\ln{\prod\limits_{m=1}^{R}f_m})=
{(\prod\limits_{m=1}^{R}f_m)}^{-1}\frac{d}{dt}(\prod\limits_{m=1}^{R}{f_m})=0$.
So, $\prod\limits_{m=1}^R{f_m}=const$. In an equilibrium state we
have $\prod\limits_{m=1}^R{f_m}=f_R$. Because the equality
$\sum\limits_{m=1}^{R}F_m=0$ is fulfilled during all time, we have
that equality, $\prod\limits_{m=1}^R{f_m}=f_R$, is a motion
integral. It is in agreement with Liouville theorem about
conservation of phase space.

So, only in two cases the Liouville equation for the whole
non-equilibrium system is in agreement with the general Liouville
equation for selected subsystems: if the condition
$\int\limits_{0}^{t}{(\sum\limits_{k=1}^L\frac{\partial}
{\partial{p_k}}F_{k}^m)}dt\rightarrow{const}$ is satisfied when
$t\rightarrow\infty$, or when,
${(\sum\limits_{k=1}^L}\frac{\partial}{\partial{p_k}}F_{k}^m)$, is
a periodic function of time. The first case corresponds to the
irreversible dynamics, and the second case corresponds to
reversible dynamics.

Thus the irreversible dynamics is possible under condition of
redistribution of phase-space volume between subsystems when full
volume is invariant. Reversibility exists, when the system is
placed near to equilibrium or in a point of invariant set of
phase-space. In the latter case a periodic change of phase-space
volume of subsystems have a place under condition of system
phase-space volume preservation as a whole.

Thus, generalized Liouville equation allows describing dynamics of
nonequilibrium systems within the framework of classical
mechanics. According to this equation both reversible and
irreversible dynamics have a place. Irreversibility is possible
only at presence of subsystems motions and dependence of the
generalized forces from subsystems velocities. Presence of such
dependence eliminate an interdiction on irreversibility, which
dictated by the Poincare's theorem of reversibility. Therefore
first of all for the proof of existence of irreversibility it is
necessary to show presence of the relative motion of subsystems in
nonequilibrium systems.

\section{The relative motion of subsystems}

The proof of existence of relative motion of subsystems in
nonequilibrium systems is following from [2] (see, $\S 10$). In
agreement with [2] the entropy $S$ for system can be writing as:
\begin{equation}
S={\sum\limits_{m=1}^R}S_m(E_m-T_m^{tr})\label{eqn2}
\end{equation}
Here, $S_m$ is entropy for $m$-subsystem; $E_m$ is full energy of
the $m$-subsystem; $T_m^{tr}=P_m^2/2M_m$ is a kinetic energy of
motion of $m$-subsystem; $M_m$ is a subsystems mass; $P_m$ is a
momentum. The argument in $S_m$ is internal energy of
$m$-subsystem.

As the system is closed, we have:
${\sum\limits_{m=1}^R}P_m=const$,
${\sum\limits_{m=1}^R}{[r_mP_m]}=const$. Here $r_m$ is a position
vector of $m$-subsystem.

The entropy in equilibrium state as a function of momentum of
subsystems has a maximum. Using a method of uncertain Lagrange
multipliers, it is possible to determine necessary conditions of a
maximum if to equate to zero the derivatives with respect to
momentum from the following expression:
\begin{equation}
{\sum\limits_{m=1}^R}\{S_m+aP_m+b[r_mP_m]\},\label{eqn3}
\end{equation}
where $a, b$ are constant multipliers.

Differentiating $S_m$ with respect $P_m$, taking into account
definition of temperature, we shall obtain: $\frac{\partial}
{\partial{P_m}}S_m(E_m-T_M^{tr})=-P_m/(M_mT)=-v_m/T$. Hence,
differentiating eq.(3) with respect $P_m$, we shall have:
$v_m=u+[{\Omega}r_m]$ (a), where ${\Omega=bT}, u=aT$, $T$ is a
temperature. From here follows, that the entropy has a maximum,
when velocities of all subsystems are determined by the formula
(a). According to this formula in equilibrium all subsystems
should to move with identical translational velocities and to
rotate with identical angular velocity. It means, that a closed
system in equilibrium state can only move and rotate as the whole;
any relative motions of subsystems are impossible.

As it follows from the formula (2), the rate of systems deviation
from equilibrium is determined by value $T_m^{tr}$. This energy
can be selected by dividing the system on equilibrium subsystems.
At such splitting all energy of system strictly equals the sum of
two types of energy. The first type of energy is the sum of
internal energy of subsystems. The second type of energy is
$T_m^{tr}$. In connection with eq. (2), the process of
equilibration is caused by transformation of energy, $T_m^{tr}$ in
internal energy of systems.

Thus, equilibrium is characterized by a condition, $T_m^{tr} =0$,
which have a place at any splitting of the equilibrium system into
subsystems. Otherwise a subsystems will have the relative motion.
Therefore a subsystems in nonequilibrium system will have relative
motion. If the system goes to equilibrium, the energy $T_m^{tr}$
should aspire to zero also.

Below, using a method of splitting into subsystems, and basing on
generalized Liouville equation, we shall view the closed
nonequilibrium a hard-disk system. We shall show, that in such
system the energy, $T_m^{tr}$, is transformed by irreversible way
into internal energy as a result of the generalized forces work
[18].

\section{Irreversibility for a hard-disks system}

The equation of motion for hard disks is deduced on the basis of a
matrix of collision from laws of conservation of energy and a
momentum. This equation can be written so [20]:
\begin{equation}
\dot{V}_k=\Phi_{kj}\delta (\psi_{kj}(t))\Delta_{kj}\label{eqn4}
\end{equation}
where $\Phi_{kj}=i(l_{kj}\Delta_{kj})/(|l_{kj}||\Delta_{kj}|)$;
$\psi_{kj}=[1-|l_{kj}|]/|\Delta_{kj}|$; $\delta(\psi_{kj})$-delta
function; $l_{kj}(t)=z_{kj}^0+\int \limits_{0}^{t}\Delta_{kj}{dt}$
are distances between centers of colliding disks;
$z_{kj}^0=z_k^0-z_j^0$, $z_k^0$ and $ z_j^0$ - are initial values
of disks coordinates; $k$ and $j$ are numbers of colliding disks;
$i$ is an imaginary unit; $t$ is a time; $z_{kj}^0=z_k^0-z_j^0$
are initial value of disks coordinates; $\Delta_{kj}=V_k-V_j$ are
relative disks velocities; $D$ is a diameter of disks. The
collisions are considered to be central, and friction is
neglected. Masses and diameters of disks are accepted to be equal
to 1. The moments of collisions $k$ and $j$ disks are determined
by equality $\psi_{kj}=0$. The collisions are considered to be
central, and friction is neglected. Masses and diameters of disks
are accepted to be equal to 1.

The eq. (4) is a non-Newtonian equation because the forces depend
on relative disks velocities. As the disks are absolutely rigid,
the internal degrees of freedom in them do not exist. Therefore
introduction of potential energy is contradicting the condition of
rigidity of disks. But potential energy can be used a formally if
determined it by delta of function. The eq. (4) is shown, that the
disks dynamics is determine by the redistribution of kinetic
energy without its transformation to potential energy. As the
force of interaction of the disks depends on their relative
velocities, the generalized subsystems forces will depend also on
subsystems velocities.

Now let us consider the question, how equilibrium is established
by the mixing. Let us take the nonequilibrium system of disks
consisting from two equilibrium subsystems: $L$ and $K$
consequently. Let us  $L$- subsystem will fly on $K$- subsystem.
Let's assume that all disks collide simultaneously through equal,
short enough intervals of time $\tau$. Such assumption does not
influence qualitative characteristics of evolution. Then the
equation of motion of disks (4) will become [17]:
\begin{equation}
\dot{V}_k^n=\Phi_{kj}^n\Delta_{kj}^{n-1}\label{eqn5}
\end{equation}

Here $k$ is a disks from $L$-subsystem. To each $k$ and the moment
of time $t=n{\tau}$ there corresponds number $j$;
${{k}{\neq{j}}}$.

The evolution of the subsystem is determined by the vector-column
$\vec{V_L}$, which components is a velocities of disks of the $L$-
subsystem; $\vec{V_L}=\{V_k\}, k=1, 2, 3,...L$. Some of the
evolution's properties of this subsystem will be determined by
behavior with the time of the sum its components. Let us designate
this sum as $V_L$. Carrying out the summation in (5) on all disks
of the subsystem, we shall obtain [17]:
\begin{equation}
\dot{V}_L^n=
\sum\limits_{k=1}^L\Phi_{ks}^n\Delta_{ks}^{n-1}\label{eqn6}
\end{equation}

Here we taken into account, that the contribution to the right
hand side of eq. (6) gives collisions of disks $L$-subsystem with
external $s$ - disks. This equation is written down in approach of
pair collisions.

The eq. (6) describes change of a total momentum, effecting onto
the  $L$-subsystem as a result of collisions at the moment
$n\tau$. The aspiration of a total momentum to zero is equivalent
to aspiration to zero of force, acting on the part of external
disks. Thus the eq. (6) is determining the relative subsystems
velocities.

When $L\rightarrow\infty$, from a condition of mixing the
uniformity of distribution of impact parameters of disks follow.
Really, according to definition of mixing, for system of disks
fairly a condition [14]: $\mu(\delta)/\mu(d)=\delta/d$ (b). Here,
$\mu(\delta)$ is a measure, corresponding to the total value of
impact parameter "$d$"; $\delta$ is an arbitrary interval of
impact parameters and, $\mu(\delta)$, is a corresponding measure.
The fulfillment of the (b) condition  means the proportionality
between the number of collisions of disks, falling at the
interval, $\delta$, and the disk diameter-$d$. I.e. the
distribution of impact parameters is homogeneous.

At performance (b) the condition of decay correlations is fair.
Therefore from eq. (6) follows:
$<\Phi_{ks}^n\Delta_{ks}^{n-1}>=<\Phi_{ks}^n><\Delta_{ks}^{n-1}>$.
As the first multiplier depends on impact parameters, and the
second depends on relative velocities, the condition of decay of
correlations is equivalent to a condition of independence of
coordinates and momentums [3, 14, 15]. I.e., when
$L\rightarrow\infty$, it is possible to pass from summation to
integration phase of the multiplier $\phi=<\Phi_{ks}^n>$ on impact
parameter. Then we will have [17]:
$\phi=\frac{1}{L}\lim\limits_{L\rightarrow\infty}\
\sum\limits_{k=1}^L\Phi_{ks}^n=
\frac{1}{G}\int\limits_0^{\pi}\Phi_{ks}^nd(\cos\vartheta)=-\frac{2}{3},$
where $G=2$ is normalization factor; $(\cos\vartheta)=d$ is impact
parameters.

As their velocities of the center of mass at determine relative
velocities of subsystems, we shall have:
$V_L^n=<\Delta_{ks}^{n-1}>$. Therefore, we have:
$V_L^n=-\frac{2}{3}V_L^{n-1}$.

Thus, the velocity of a subsystem is decrease. The rate of
decreasing is determined by factor 2/3. In a result the system is
equilibrates. In our case the sense of replacement of summation on
integration on impact parameters will consist only in transition
from the discrete description to continuous one. It has allowed
estimating the rate of an establishment of an equilibrium state.
It will be clear from below, this summation has no relation to the
process of the irreversibility proving, as in "coarse-grain" case
it have a place [17].

Let us designate the equilibrium point as $Z_0$. As it followed
from eq. (6), this point is asymptotically steady [17]. Stability
is provided with occurrence of returning force $F_L$ (right hand
side of the eq. (6)) acting on $L$-subsystem  at a deviation of it
from an equilibrium point. Therefore the stability of equilibrium
state leads to limitation of permissible amplitudes of
fluctuations of the system parameters. Really, any nonequilibrium
condition is characterized by force $F_L$. So we have:
$\dot{V}_L=F_L$. As a result of mixing this force should decrease.
Time of decrease of force is determined from a condition
$t=\int\frac{dV_L}{F_L}$. Hence, if the system in the artificial
way appeared in a nonequilibrium point, through characteristic
time, $t_{din}\sim\frac{1}{F_L}$, this system will come to
balance. Hence, the rate of nonequilibrium is determined by the
value $F_L$. So, are possible only those fluctuations, for which
condition, $t_{fluct}<t_{din}$ is satisfied. The time $t_{fluct}$
is determined probabilistic laws. According to the formula
Smoluhovsky [15], for a case ergodic of systems average resetting
time, $t_p$, or Poincare's cycle times is equal
$\tau=t_p(1-P_0)/(P_0-P_1)$, were $P_1$ -is a probability of
reversibility during the time $t_p$; $P_0$ is a probability of
initial phase region. So, the originating of forces $F_L$ at
spontaneous deviation of a system from equilibrium, leads to
limitation of permissible amplitudes of fluctuations. Only such
spontaneous fluctuations are possible, which one does not
contradict a condition $t_{din}>\tau$.

Thus, spontaneous motion of system from equilibrium point is
possible only into such points of phase-space, for which the
characteristic time of return, determined to corresponding these
points by a field of forces, more than the characteristic time
necessary for a spontaneous deviation. From here follows, that
framework of applicability probabilistic descriptions of dynamics
of systems, and, also, the rate of possible fluctuations, are
determined by a condition $t_{prob}<t_{din}$ [17].

Though the mechanism of irreversibility offered here bases on
property of mixing, the coarse-grain problem  is eliminated in it.
Really, in a basis of this mechanism the dependence of the
generalized force on velocities of elements is lays. Evolution of
such systems is described by the generalized Liouville equation.
The generalized Liouville equation, as against the canonical
prototype, does not forbid the existence of irreversible dynamics.

For a hard disks system we have the following explanation of the
mechanism of irreversibility [18, 21]. Subsystems of disks in
nonequilibrium systems possess relative motion. As a result of
their interaction, because of mixing property of phase-space, the
chaotic state of disks velocities is increases. It is lead to
transformation of energy $T_m^{tr}$ into internal energy,
reduction of the relative subsystems velocities and their
interaction forces. The process of increasing of the relative
velocities of subsystems is forbidden due to the law of
preservation of a momentum (it will be discussed below more
carefully). Therefore the system equilibrates.

Thus, the proof of existence of irreversible dynamics of hard
disks is based on the dependence of forces of interaction of disks
on their velocities. But all forces in the nature are potential
[22]. And according to the eq. (1) for existence of
irreversibility in systems of potentially interacting elements,
the presence of dependence of the generalized forces from
velocities of motion of subsystems is necessary. It will be shown
below, that such dependence takes place in nonequilibrium systems
of potentially interacting elements.

\section{The subsystems dynamics}

In this section presence of dependence of the generalized forces
from velocities in nonequilibrium systems of potentially
interacting elements will be shown. With this purpose we shall
obtain the equation, describing an energy exchange between
subsystems. Basing on it we shall find an analytical form of the
generalized forces.

Let us take the system consisting from $N$ elements. Masses of
elements are accepted to $1$. We shall present energy of the
system as the sum of kinetic energy of the motion of system as the
whole-$T_N^{tr}$; the kinetic energy of the motion of its elements
concerning the center of mass- $\widetilde{T}_N^{ins}$; and their
potential energy- $\widetilde{U}_N^{ins}$. The energy,
$E_N^{ins}=\widetilde{T}_N^{ins}+\widetilde{U}_N^{ins}$, is
internal energy of the system. It is the sum of the kinetic energy
of the relative motion of elements and the energy of the potential
interaction. Relative elements velocities and distances between
them determine the internal energy.

The energy, $T_N^{tr}$ is determined by the velocity, $V_N$ of the
center of mass. This energy is depended on the rate of regularity
of the particles velocities because
$V_N=\frac{1}{N}\sum\limits_{i=1}^Nv_i$ .

When the external forces are absent, the energies, $T_N^{tr}$ and
$E_N^{ins}$ are constants. It is because in connection with the
momentum preservation law these energies are the motion integrals.

The full energy of the closed system of potentially interacting
elements in homogeneous space can be presented so:
$E_N=T_N+U_N=const$, where
$T_N=\frac{1}{2}\sum\limits_{i=1}^N{{v_i}^2}$ is a kinetic energy;
$U_N(r_{ij})$ is potential energy; $r_{ij}=r_i-r_j$ is the
distance between $i$ and $j$ elements.

The equation of motion for elements of system can be obtained
differentiating the expression of energy systems with respect to
time [7, 12, and 19]. We will have:
$\dot{v}_i=-\sum\limits_{i=1,j\neq{i}}^N\frac{\partial}
{\partial{r_{ij}}}U$ (c). The eq. (c) generally is nonlinear. Let
us consider the nature of this nonlinearity on the example of
two-body system. In the laboratory system of coordinates the
kinetic energy of particles can be divided into energy of motion
of the center of mass of system, $T^{tr}$, and energy of the
particles motion relative to the center of mass, $T_i^{ins}$,
where $i$ is a number of subsystems (in this case "i" is number of
elements, $i=1, 2$). Eventually the redistribution character  of
these types of energies between particles is various. So, the
change of energy $T_i^{ins}$  connected with it transformation
into the particles potential energy. The energy,
$T^{tr}=T_1^{tr}+T_2^{tr}=const$ is redistributed between
particles by nonlinear way without its transformation to potential
energy. By transition to the system of coordinates of the center
of mass, the nonlinearity and non-potentiality for two-body system
are eliminated because this operation is excludes kinetic energy
of motion of the center of mass. In a result the task becomes
integrable. For the systems of three and more bodies the excluding
of nonlinearity by this way in general case is impossible.
Therefore these systems are not integrable.

From here becomes clear why it is necessary to divide system into
the equilibrium subsystems. By such splitting the nonlinearity of
dynamics caused by relative motion of microsystems inside
subsystems is excluded. If the subsystem is in equilibrium, it
does not matter how we divide it on the microsystems. In any case
these microsystems will be motionless relative to each other and
nonlinearity of dynamics is absent. Owing to splitting of
nonequilibrium system into equilibrium subsystems, the question
about irreversibility nature is reduced to a problem about
character of subsystems energy exchange. To emphasize the absence
of the energy relative motion of the microsystems in internal
energy of an equilibrium subsystem, we shall name internal energy
of equilibrium subsystem as "bound energy".

Let us to take the system consisting of two interacting
equilibrium subsystems. It is $L$, and $K$-subsystems. The $V_L$
and $V_K$ are velocities of the center of mass of corresponding
subsystems. The number of elements in  $L$-subsystem is equal to
$L$, and the number of elements in $K$ -subsystem is equal to $K$.
Let us equalities, $L+K=N$ and $LV_L+KV_K=0$, have a place, i.e.
the center of mass of system is motionless.

It is obvious, if the interaction of subsystems will strong
enough; these subsystems can be broken on the different number of
smaller equilibrium microsystems. It will complicate the analysis
though will not bring in qualitative changes to process of
evolution. Therefore we shall accept that interaction of
subsystems is weak enough. It will allow accepting that the
subsystems are in equilibrium during all process of interaction.

The equations for energy exchange between subsystems can be
obtained, by differentiating system energy with respect of time,
grouping together the terms, which corresponds to elements
different subsystems. Having executed necessary transformations
and having allocated the terms corresponding to the different two
types of energy: the bound energy and energy of motion of
subsystems, the next equations we shall obtain [21]:
\begin{equation}
{LV_L\dot{V}_L+{\sum\limits_{j=i+1}^L}\sum\limits_{i=1}^{L-1}\{v_{ij}
[\frac{\dot{v}_{ij}}{L}+\frac{\partial{U}}{\partial{r_{ij}}}]\}=
-\sum\limits_{{j_K}=1}^K}\sum\limits_{{i_L}=1}^{L}v_{i_L}
\frac{\partial{U}}{\partial{r_{{i}_{L}{j}_{K}}}} \label{eqn7}
\end{equation}

\begin{equation}
{KV_K\dot{V}_K+{\sum\limits_{j=i+1}^K}\sum\limits_{i=L+1}^{K-1}\{v_{ij}
[\frac{\dot{v}_{ij}}{K}+\frac{\partial{U}}{\partial{r_{ij}}}]\}=
-\sum\limits_{{j_K}=1}^K}\sum\limits_{{i_L}=1}^{L}v_{j_K}
\frac{\partial{U}}{\partial{r_{{i}_{L}{j}_{K}}}} \label{eqn8}
\end{equation}

Here, ${v_{ij}=v_i-v_j}$ are the relative velocities. The
subindexes, $L, K$, denote to which subsystems some elements
belong.

The left hand sides in the eqs. (7, 8) are determining the changes
of the energies, $T_N^{tr}$ and $E_N^{ins}$, subsystems as a
result of their interaction. The first term is set the change of
kinetic energy of motion of subsystems as the whole. The second
term describes transformation of the bound energy. The right hand
sides of the eqs. (7, 8) describe the interaction of subsystems
and determine the rate of an exchange of energy between
subsystems.

Velocities of any particles of a subsystem can be presented, as
the sum of velocities of the center of mass of a subsystem plus
their velocities concerning the center of mass. I.e.,
$v_i=\tilde{v_i}+V$. Then, having grouped both parts of the eq.
(5) in appropriate way, we shall obtain:
\begin{equation}
{LV_l{[\dot{V}_L+\sum\limits_{{j_K}=1}^K}\sum\limits_{{i_L}=1}^{L}
\frac{\partial{U}}{\partial{r_{{i}_{L}{j}_{K}}}}] +
{\sum\limits_{j=i+1}^L}\sum\limits_{i=1}^{L-1}\{v_{ij}
[\frac{\dot{v}_{ij}}{L}+\frac{\partial{U}}{\partial{r_{ij}}}]\} =
-\sum\limits_{{j_K}=1}^K}\sum\limits_{{i_L}=1}^{L}
\frac{\partial{U}}{\partial{r_{{i}_{L}{j}_{K}}}}{\tilde{v}_{i_L}}
\label{eqn9}
\end{equation}

The eq. (9) determines the change of energy of the  $L$-subsystem
at interaction it with a  $K$-subsystem. As it follows from the
right hand side term of the eq. (9), the change of energy of
$L$-subsystem as a result of its interaction with a  $K$-subsystem
is determined by velocities of motion of particles  $L$-subsystem
concerning its center of mass and potential interaction with
particles of the  $K$-subsystem.

The first term of the left hand side of eq. (9) determines the
change of kinetic energy of motion of  $L$-subsystem as a result
of its motion in a field of  $K$-subsystem. The second term
determines the change of the bound energy of a subsystem as a
result of motion of its particles in a field of $K$-subsystem
particles.

When $\dot{V_L}=0$, the energy of relative motion of subsystems is
absent, and the right hand side of the eq. (9) is equal to zero.
In this case the full system energy is equal to the sum of the
bound subsystems energies.

If the forces of particles interaction inside subsystems will be
infinity, the velocities of motion of particles inside subsystems
can be neglected. It corresponds to equality to zero of the right
hand side of eq. (9). In this case the second term in the left
hand side of eq.(9) is equal to zero, and this equation will
transformed to the usual equation of Newton describing motion of
two hard bodies.

\section{Difference between particles and subsystems dynamics}

The Newton equations (c) can be treated, as equation for the
particles interaction forces. The work of these forces is
determining transformation of kinetic energy of particles to their
potential energy. This energy transformation occurs at transition
of system from one point of configuration space into another [12].
Forces are set by a gradient of the potential energy of particles.
Thus, the forces and potential energy of particles are completely
determined by coordinates, and work of potential forces on the
closed contour is equal to zero. It corresponds to reversible
dynamics.

And now we shall consider the eq. (9). From it follows, that in
nonequilibrium systems the kinetic energy of relative motion of
subsystems is appeared. This energy is connected with the rate of
regularity of particles motion of subsystems. The regularity is
determined by deviation  from equilibrium of the velocities
distribution functions. As against Newton's forces, the work of
the generalized forces between subsystems will transform kinetic
energy of motion of subsystem not only to the potential energy of
a subsystem as the whole, but also into the bound energy. Because
of such transformation, the work of the generalized forces on the
closed contour in configuration space is distinct from zero.

Thus, transformation of kinetic energy of relative motion of
subsystems into the bound energy is determined by the work of the
generalized forces. As a result the kinetic energy of relative
motion of subsystems disappears. But in this process the kinetic
energy of particles is not obliged to change. Really, in the
coordinates of the center of mass of system, it is possible to see
that as a result of interaction of subsystems the orderliness of
velocities of particles decreases. But in the coordinates of the
centers of mass of each subsystem we will find a decreasing of
energy of motion and relative velocities of subsystems, and
increasing of their bound energy. This process is caused by the
increasing of directions disordering of the velocities vectors of
particles in a result of work of the subsystems interaction force.
And the system is equilibrates.

The transition of the bound energy into kinetic energy of a
subsystem is impossible. It is a cause of irreversibility. Really,
this transition would be possible only under condition of
spontaneous occurrence inside an equilibrium subsystem of the
generalized forces. But their occurrence would mean infringement
of spherical symmetry of function of distribution of velocities of
elements of equilibrium subsystems concerning the center of mass.
And it contradicts the law of preservation of a momentum.

Thus, the eq. (9) as against the Newton equation describes process
of transformation of energy in the system, caused not only by
transformation of the potential energy into kinetic energy, but
also by change of function of distribution of velocities of
particles due to increasing the rate of chaotic motion of the
particles.

There is a question why the Newton equation fairly for the
description of dynamics of particles, but, nevertheless, it does
not determine system equilibration? Let us to offer the following
answer. Dynamics of selected particles is unequivocally determined
by the equation of Newton. A motion of any particle is reversible.
But the collective parameters describing subsystem dynamics such
as a bound energy, the generalized forces, ambiguously depend on
the particles motion parameters. Such ambiguity leads to
occurrence of new legitimacies of systems which are not proper for
separate particles. Let us view, for example, velocity of a motion
of centre of masses of system. It is maintained in the homogeneous
space. This velocity is collective parameter of system and is
determined by the total of velocities of all particles of system
that leads to lack  biunique conformity between velocity of center
of mass and particles velocities. The impossibility of
magnification of energy of a motion of an equilibrium subsystem
due to its bound energy is the collective legitimacy determining
its dynamics. Therefore, despite of reversibility of dynamics of a
separate particle, dynamics of their subsystems can be
irreversible. Thus, irreversibility is a new property of systems
which absent in dynamics of their selected particles. Occurrence
of this property within the limits of laws of classical mechanics
becomes possible, owing to ambiguous dependence of parameters of
collective of particles on the parameters determining dynamics of
selected particles.

Having excluded from the eq. (9) the potential interaction, we
shall obtain the equation for the elastic disks. It means that
both in systems of elastic disks, and in systems of potentially
interacting elements, the nature of irreversibility are identical.

\section{Classical mechanics and thermodynamics}

It is possible to come to thermodynamics with the help of the eqs.
(7, 8). Really, the right hand side of these equations determines
an exchange of energy between subsystems as a result of their
interaction. The first term of the left hand side of each equation
determines the change of the motion energy of subsystem as the
whole. In thermodynamics it corresponds to mechanical work, which
is carried out by external forces acting on subsystem on the part
of an environment. The second term of the left hand side
corresponds to increase in the bound energy of a subsystem due to
energy of relative motion of subsystems. In thermodynamics this
term corresponds to the change of thermal energy of system.

It is easy to see the relation between the eq. (7) and the basic
equation of thermodynamics [2, 3]:${dE=dQ-{PdY}}$. Here, according
to common terminology, $E$ is internal energy of a subsystem; $Q$
is thermal energy; $P$ is pressure; $Y$ is volume.

The energy change of the selected subsystem is due to the work
made by external forces. Therefore, the change in full energy of a
subsystem corresponds to $dE$.

The change of kinetic energy of motion of a subsystem as the
whole, $dT^{tr}$, corresponds to the term ${PdY}$. Really,
${dT^{tr}=VdV=V\dot{V}dt=\dot{V}dr=PdY}$

Let us determine, what term in eq. (11) corresponds to the change
of the binding energy in a subsystem. As follows from virial
theorem [6], if the potential energy is a homogeneous function of
second order of the radiuses-vectors, then
${\bar{E}^{ins}=2\bar{\tilde{T}}^{ins}=2\bar{\tilde{U}}^{ins}}$.
The line denotes the time average. Earlier we obtained that the
binding energy, ${E^{ins}}$, increases due to contribution of
energy, ${T^{tr}}$. But the opposite process is impossible.
Therefore the change of the term $Q$ in the eq. (11)  corresponds
to the change of the binding energy ${E^{ins}}$.

Let us consider the system near to equilibrium. If the subsystem
consist of ${N_m}$ elements, the average energy of each element
becomes, ${\bar{E}^{ins}={E}^{ins}/N_m=\kappa{T}_0^{ins}}$. Now
let the binding energy increases with ${dQ}$. According to the
virial theorem, keeping the terms of the first order, we have:

${dQ\approx{T}_0^{ins}[d{E}^{ins}/{T}_0^{ins}]
={T}_0^{ins}[{dv}/{v_0}]}$, where ${v_0}$ is the average velocity
of an element, and ${dv}$ is its change. For subsystems in
equilibrium, we have ${dv/v_0\sim{{d\Gamma_m}/{\Gamma_m}}}$, where
${\Gamma_m}$ is the phase volume of a subsystem, ${d\Gamma_m}$
will increase due to increasing of the subsystem energy on the
value, ${dQ}$. By keeping the terms of the first order we get:
${dQ\approx{T}_0^{ins}d\Gamma_m/\Gamma_m={{T}_0^{ins}}d\ln{\Gamma_m}}$.
By definition ${d\ln{\Gamma_m}=dS^{ins}}$, where ${S^{ins}}$ is a
subsystem entropy [2, 3]. So, near equilibrium we have
${dQ\approx{T}_0^{ins}dS^{ins}}$.

\section{Relation of the generalized forces with entropy}

Let us consider the relation of the generalized forces with
entropy. According with the formula (2) the entropy production in
the non-equilibrium system is determined by transformation of the
kinetic energy of subsystems motion into the bound energy.
Eventually relative velocities of subsystems and the generalized
forces go to zero. In result the energy of relative motion of
subsystems completely transforms into the bound energy and the
systems equilibrates. It means that energy of motion of a
subsystem goes on increase of entropy. Therefore the deviation of
entropy from equilibrium is determined by the next formula [21]:
\begin{equation}
{{\Delta{S}}={\sum\limits_{l=1}^R{\{{m_l}
\sum\limits_{k=1}^{m_l}\int{\sum\limits_s{{\frac{{F_{ks}}^{m_l}v_k}{E^{m_l}}}}{dt}}\}}}}\label{eqn10}
\end{equation}

Here ${E^{m_l}}$ is the kinetic energy of subsystem; ${m_l}$ is
the number elements in subsystem ${"l"}$; ${R}$ is the number of
subsystems; ${s}$ is number of the external disks which collided
with internal disk ${k}$; ${F_{ks}^{m_l}}$ is a force, acted on
disk $k$-disks; $v_k$ -velocity of the $k$- disk.

The integral (10)is determining the work of the force,
${F_{ks}^{m_l}}$ during the system relaxation to equilibrium. In
equilibrium the energy of the relative motion of subsystems and
generalized forces are equal to zero. I.e. the integral in eq.
(10) is determined by the energy of relative motion of subsystems.
It is corresponds to phenomenological formula Clauses for entropy
[3]. So the eq. (10) will be in agreement with the eq. (5).

Really, if ${E_l^{ins}\gg{T_l^{tr}}}$, than we have:
${dS=\sum\limits_{l=1}^R\frac{\partial{S_l}}{\partial{T_l^{tr}}}{dT_l^{tr}}}$.
It is corresponds to eq. (10). Both in eq. (5) and in eq. (10)
entropy increasing is determined by change the energy of the
relative motion of subsystems.

Thus, the eq. (10) connects dynamic parameter - force acting on a
subsystem, with entropy which is a thermodynamic parameter. I.e.
eq.(10) establishes connection between parameters of classical
mechanics and thermodynamic parameters. The deviation of system
from equilibrium is characterized by the ratio between energy of
relative motion of subsystems and full energy of system.

Thus the interrelation of Boltzmann's entropy definition, which
based on the measure of chaos, and definition of the Clauses
entropy is cleared by eq. (10).

\section{Discussion}

At present time the problem of irreversibility was reduced to the
problem of the nature of coarse-grain of the phase-space. All
attempts to solve this problem within the framework of classical
mechanics encountered to Poincare's theorem of reversibility. The
proof of this theorem is based on a strict formalism of Hamilton,
in particular, on canonical Liouville equation. The impossibility
of "coarse-grain" follows from this equation. It seems that this
fact deprives of any hopes for successful solving the problem of
irreversibility [10, 23]. But nevertheless we found this solution
by expansion of a canonical formalism to open systems. This
solution is in the framework of the laws of classical mechanics
but not in the framework of canonical Hamilton's formalism.

Already during solving a problem of three bodies doubts appeared
concerning completeness of methods of classical mechanics. Somehow
or other the difficulty of its solution is connected to a problem
of the description of nonlinear process of an exchange of energy
between bodies. In statistical physics this problem compelled to
be limited to make researches of only equilibrium systems for
which it was possible to neglect an exchange of energy between
subsystems. In kinetics the account of streams of energy,
substance, etc. was based on phenomenological formulas [2-6]. By
this way a problem of the description of character of energy
exchanging between interacting systems within the framework of
classical mechanics has been bypassed. But the knowledge of this
mechanism is necessary for understanding of the equilibration
process nature. From here follows that for solving a problem of
irreversibility it is necessary to find a method of describing of
process of energy exchange between interacting systems. It
explains our aspiration to expand the Hamilton formalism so that
it would be possible to describe dynamics of opens systems in its
framework.

The expansion of the formalism and research of the mechanism of
irreversibility were carried out by us simultaneously. As a models
the nonequilibrium many-body system were used. The system was
divided into the equilibrium subsystems. To describe character of
an energy exchange, we found the equation in which the energy of
system is represented as the sum of energies of subsystems
motions, their bound energy and interaction energy. Such
representation of energy has played a main role in disclosing of
the irreversibility mechanism.

At the beginning of our investigation the colliding hard disks
were studied [20]. Their interaction is determined by a matrix of
collisions. With its help the equation of motion of hard disks has
been obtained. This equation turned out to be non-Newtonian
because during collisions of disks there is a redistribution of
kinetic energy between them without it transformation into the
potential energy. I.e. the use of concept of potential energy for
the description of hard disks dynamics turned out to be
superfluous. From here there was a necessity for searching such
expansion of Hamilton formalism which will allow to describe not
potentially of systems interacting. Such expansion was created in
a framework of classical mechanics by variational methods basing
on D'Alambert equation [16, 17]. Expansion is consisted in
replacement of canonical Lagrange, Hamilton and Liouville
equations to the corresponding generalized equations.

The generalized Liouville equation obtained by us is applicable
for the description of any systems, as potentially interacting
systems so non-potentially. From this equation follows that the
dependence of subsystems interaction forces on velocities is
necessary for existence of irreversibility. Taking into account
the mixing property and presence of dependence of forces between
disks from velocities with the help of generalized Liouville
equation an opportunity of existence, both reversible and
irreversible dynamics was proved. But studying of disks dynamics
was a preliminary step on a way to understanding of the
irreversibility nature. Really, non-Newtonian forces determine
dynamics of disks, but in real systems fundamental forces are
potential [21]. Therefore the generalization these results on
systems of potentially interacting elements were required.

The disks researches have shown that the question about
irreversibility is reduced to a problem of existence of subsystems
velocities and dependence on velocities of their interacting
forces. Presence of such dependence in non-equilibrium systems was
proved in [2]. For equilibrium systems the subsystems motions are
absent. Therefore it was necessary to know, whether irreversible
transformation of kinetic energy of relative motions of
potentially interacting elements subsystems to other types of
energy is possible. The Newton equation did not suit for this
purpose. Really, it describes only such transformations of energy,
which are connected to reversible transition of kinetic energy
into potential and on the contrary. On the other hand, the Newton
equation is fairly convenient for the description of dynamics of
elements of any systems in which energy is kept. Therefore we
assumed that the Newton equation is not responsible for
transformation subsystems relative motion energy. For checking it
up, it was necessary to obtain the equation of an energy exchange
between subsystems. We have obtained this equation from the law of
conservation of energy. From this equation the expression for the
generalized force of interaction of subsystems has been found.

According to the equation of energy exchange between subsystems,
energy of their relative motion will be transformed both in their
potential and into the bound energy. And energy transformation to
the bound energy occurs as a result of increase of particles
motion randomness. Such transformation of energy conducts to
irreversible decrease of energy of motion of a subsystem.
Irreversibility is provided with the law of preservation of a
momentum of subsystems. From here the next explanation of the
mechanism of irreversible dynamics follows.

The rate of the systems nonequilibrium is determined by the
organizing of motion of systems elements. Therefore if we divide
nonequilibrium system into the equilibrium subsystems, the
relative motion of subsystems should exist due to the organizing
of elements. As a result of the work of the generalized forces the
energy of relative motion of a subsystem decreases. This energy is
transformed both into the potentially and into the bound energy of
subsystems. The process of increase of the bound energy is going
due to increasing of a randomness of vectors of subsystem elements
velocities. The mixing is a cause of the increase of randomness.
Process of reduction of subsystems relative motion energy is
irreversible because of the impossibility of increasing of their
relative velocities due to the bound energy. It is follows from
the law of preservation of a momentum of subsystems. Equilibrium
is established when the subsystems relative motion kinetic energy
completely transformed into the bound energy.

Interaction of elements of systems with each other is essence for
the offered mechanism of irreversibility occurrence. Therefore
this mechanism is unsuitable for ideal gas or Brownian particles
because for them the establishment of equilibrium is determined by
interaction with external environment which set in the stochastic
way [24, 25]. Probabilistic laws are describing the process of
establishment of equilibrium state in these systems.

The offered mechanism of irreversibility cannot be obtained on the
basis of the canonical formalism equations for closed systems, as
these equations do not describe the work of non-potential forces
of interaction of the subsystems reducing orderliness of particles
motion. Though this mechanism is connected to property of mixing,
the problem of coarse-grain phase-space here does not arise.
Moreover the coarse graining of the phase-space is sequent from
this mechanism because it corresponds to the system equilibration
when integrated velocity of particles in any physically small
volume should aspire to zero.

The mechanism of irreversibility submitted here determines the
connection between classical mechanics and thermodynamics. Really,
the first law of thermodynamics follows from the equation
describing energy transformation of interacting subsystems. This
equation is determined by presence of two qualitatively various
types of energy: the bound energy and kinetic energy of relative
motion of a subsystem as the whole. Irreversible transition of
subsystems motion energy into their bound energy as a result of
increase of randomness determines the contents of the second law
of thermodynamics. Really, energy of subsystems motion as a result
of chaos increase goes to entropy increase.

The further development of subsystems dynamics researches on the
basis of the offered approach represents significant interest.
These researches will help to prove thermodynamic laws. They are
also perspective from the point of view of classical mechanics
expanded formalism creation, allowing to study systems
interactions processes and also opening systems.

\smallskip

\end{document}